\documentclass[]{aa}
\usepackage{graphicx,txfonts,amssymb,natbib,epsfig}
\titlerunning{Comparison of simulated and real strong lensing clusters}
\authorrunning{M. Meneghetti et al.}

\begin{document}

\title{Comparison of an X-ray selected sample of massive lensing clusters with the {\sc MareNostrum Universe} $\Lambda$CDM simulation}


 \author{M. Meneghetti\inst{1,3}\thanks{E-mail: massimo.meneghetti@oabo.inaf.it}, C. Fedeli\inst{2,3,4}, A. Zitrin\inst{5}, M. Bartelmann\inst{6}, T. Broadhurst\inst{9,10}, S. Gottl\"ober\inst{7}, L. Moscardini\inst{2,3}, G. Yepes\inst{8}}

\institute {
  $^1$ INAF-Osservatorio Astronomico di Bologna, Via Ranzani 1, I-40127 Bologna, Italy\\
  $^2$ Dipartimento di Astronomia, Universit\`a di Bologna,
  Via Ranzani 1, I-40127 Bologna, Italy\\
  $^3$ INFN, Sezione di Bologna, Viale Berti Pichat 6/2, I-40127 Bologna, Italy\\
  $^4$ Department of Astronomy, University of Florida, Bryant Space Science Center, Gainesville, FL, 32611 \\
  $^5$ The School of Physics and Astronomy, the Raymond and Beverly Sackler Faculty of Exact Sciences, Tel Aviv University, Tel Aviv 69978, Israel \\ 
  $^6$ Zentrum f\"ur Astronomie, Institut f\"ur Theoretische Astrophysik, Albert-\"Uberle Str. 2,  69120 Heidelberg, Germany \\
  $^7$ Astrophysikalisches Institut Potsdam, An der Sternwarte 16, D-14482 Potsdam, Germany \\
  $^8$ Grupo de Astrof\'isica, Universidad Aut\'onoma de Madrid, Madrid E-28049, Spain \\
  $^9$ Department of Theoretical Physics, University of Basque Country UPV/EHU, Leioa, Spain \\
  $^{10}$ IKERBASQUE, Basque Foundation for Science, 48011, Bilbao, Spain}

\date{\emph{Astronomy \& Astrophysics, accepted for publication}}

\abstract{A long-standing problem of strong lensing by galaxy clusters regards the observed  high rate of giant gravitational arcs as compared to the predictions in the framework of the ``standard" cosmological model. This is known as  the ``arc statistics problem". Recently, few other inconsistencies between theoretical expectations and observations have been claimed which regard the large size of the Einstein rings and the high concentrations of few clusters with strong lensing features. All of these problems consistently indicate that observed galaxy clusters may be gravitational lenses stronger than expected.}{We aim at better understanding these issues by comparing the lensing properties of well defined cluster samples with those of a large set of numerically simulated objects.}{We use clusters extracted from the  MareNostrum Universe to build up mock catalogs of galaxy clusters selected through their X-ray flux. We use these objects to estimate the probability distributions of lensing cross sections, Einstein rings, and concentrations for the  sample of 12 MACS clusters at $z>0.5$ presented in Ebeling et al. (2007) and discussed in Zitrin et al. (2010).}{We find that three clusters in the MACS sample have lensing cross sections and Einstein ring sizes larger than any simulated cluster in the MareNostrum Universe. We use the lensing cross sections of both simulated and real clusters to estimate the number of giant arcs that should arise by lensing sources at $z=2$. We find that simulated clusters produce $\sim 50\%$ less arcs than observed clusters do. The medians of the distributions of the Einstein ring sizes differ by $\sim 25\%$ between simulations and observations. We estimate that, due to cluster triaxiality and orientation biases affecting the lenses with the largest cross sections, the concentrations of the individual MACS clusters inferred from the lensing analysis should be up to a factor of $\sim 2$ larger than expected from the $\Lambda$CDM model. In particular, we predict that for $\sim 20\%$ of the clusters in the MACS sample the lensing-derived concentrations should be higher than expected by more than $\sim 40\%$.}{The arc statistics, the Einstein ring, and the concentration problems in strong lensing clusters are mitigated but not solved on the basis of our analysis. Nevertheless, due to the lack of redshifts for most of the multiple image systems used for modeling the MACS clusters, the results of this work will need to be verified with additional data. The upcoming CLASH program will provide an ideal sample for extending our comparison.}


\maketitle

\section{Introduction}
\label{sect:intro}
Strong lensing is a widely used method for investigating the inner structure of galaxy clusters \citep[see][for some examples]{KO89.1,1990ApJ...350...23B,KN03.1,2005ApJ...621...53B,2006A&A...458..349C,2009MNRAS.397..341L,2010arXiv1005.0398C}. Moreover, the statistical analysis of strong lensing events in clusters was also proposed as a tool for cosmology \citep[e.g.][]{BA98.2}. The abundance of strong lensing events, such as gravitational arcs and multiple images, is expected to be higher in cosmological models where the growth of the cosmic structures is faster at earlier epochs, such as models where some dynamical dark energy starts dominating the expansion of the universe at earlier epochs \citep{BA03.1,2005MNRAS.361.1250M,ME05.1}. Thus, in these models a larger number of potential lenses populates the universe up to high redshift ($z>0.5$). The cluster concentrations are found in numerical simulations to reflect the density of the universe at the cluster epoch of formation. Clusters forming earlier have higher concentrations \citep{DO03.2} and are expected to be more efficient lenses.  

In the era precision cosmology, strong lensing statistics cannot be as competitive as other cosmological probes for constraining cosmological parameters. These are all supporting the ``concordance model"  $\Lambda$CDM model, which became the standard scenario of structure formation \citep{2009ApJS..180..330K,RI98.1,RI04.1,PE99.1,2005ApJ...633..560E,2007MNRAS.381.1053P}.  However, previous attempts of using strong lensing statistics as a cosmological tool have produced controversial results. In particular, \cite{BA98.2}, studying the lensing properties of a set of numerically simulated galaxy clusters, argued that the $\Lambda$CDM cosmological model fails at reproducing the observed abundance of giant gravitational arcs by almost an order of magnitude. This inconsistency between strong lensing and other observational data is known as the ``arc statistics problem". Its nature is still under debate. A long series of papers have tried to falsify the theoretical predictions of \cite{BA98.2}. Although several limits where found in their simulations, which could not properly capture several important features of both the lenses and the sources \citep[e.g.][]{DA03.1,WA03.1,ME03.1,TO04.1,PU05.1,ME07.1,2008ApJ...676..753W,2010MNRAS.tmp..671M}, the controversy is not yet solved. Moreover, it has been recently enforced by several other observations of strong lensing clusters, which seem to indicate that 1) some galaxy clusters have very extended Einstein rings (i.e. critical lines) {  whose abundances} can hardly be reproduced by cluster models in the framework of a $\Lambda$CDM cosmology \citep{2008MNRAS.390.1647B,2004ApJ...607..125T}, and 2) few clusters, for which high quality strong and weak lensing data became available, have concentrations  which are way too large compared to the expectations \citep{2008ApJ...685L...9B,2009MNRAS.396.1985Z}. These evidences push in the same direction of the ``arc statistics" problem, in the sense that they both suggest that observed galaxy clusters are too strong lenses compared to numerically simulated clusters.

Understanding the origin of these mismatches between theoretical predictions and observations is fundamental, as these may evidence a lack of understanding of the cluster physics, which may be not well implemented in the simulations, or, conversely, highlight some inconsistencies between the $\Lambda$CDM scenario and the properties of the universe on small scales.  
So far a comparison between theoretical predictions and observations has been complicated by the lack of systematic arc surveys but also by the fact that different approaches were used to analyze simulations and observations \citep{2008A&A...482..403M}. In this paper, we propose a novel approach, whose principal aim is to eliminate most of the assumptions used in the previous works. It consists of analyzing observed galaxy clusters, for which detailed mass models are available through strong, and possibly also weak lensing observations, fully consistently to numerical simulations. The deflection angle maps provided by the lens models are used for performing ray-tracing simulations and for lensing the same source population used in numerical simulations. We attempt a comparison between the properties of simulated clusters in the {\sc MareNostrum Universe} and those of a complete sample of X-ray luminous MACS clusters, for which strong lensing models were recently derived by \cite{2010arXiv1002.0521Z}. 

The plan of the paper is as follows. In Sects. 2 and 3 we describe the numerical and the observed cluster samples used in this study. In Sect. 4 we illustrate the analysis performed to measure the lensing cross sections and the Einstein ring sizes. In Sect. 5 we discuss the comparison between simulations and observations, describing how we construct mock cluster catalogs to simulate a MACS-like survey, and showing the statistical distributions of both cross sections and Einstein rings. Finally, we use the simulated clusters for estimating the concentration bias expected for the MACS sample. Sect. 6 is dedicated to the discussion and the conclusions.  

\section{The {\sc MareNostrum Universe}}
Our theoretical expectations are based on the analysis of the simulated clusters contained in the {\sc MareNostrum Universe}. A detailed description of the analysis performed on these objects can be found in \cite{2010arXiv1003.4544M} and in \cite{2010arXiv1007.1551F}. We briefly summarize the most relevant aspects of the analysis here. 

The {\sc MareNostrum Universe}  \citep{2007ApJ...664..117G} is a large-scale cosmological non-radiative SPH
simulation performed with the {\sc Gadget2} code \citep{SP05.1}. It was performed assuming a $\Lambda$CDM cosmological background with WMAP1 normalization, namely $\Omega_{\mathrm{m},0} = 0.3$, $\Omega_{\Lambda,0} = 0.7$ and $\sigma_8 = 0.9$ with a scale invariant primordial power spectrum. The simulation consists of a comoving box size of $500\;h^{-1}$ Mpc containing $1024^3$ dark matter particles and $1024^3$ gas particles. The mass of each dark matter particle equals $8.24 \times 10^{9} M_\odot h^{-1}$, and that of each gas particle, for which only adiabatic physics is implemented, is $1.45 \times 10^{9} M_\odot h^{-1}$.   The baryon density parameter is set to $\Omega_{\mathrm{b},0} = 0.045$. The spatial force resolution is set to an equivalent Plummer
gravitational softening of $15$~h$^{-1}$~kpc, and the SPH smoothing
length was set to the 40th neighbor to each particle.

As described in \cite{2010arXiv1003.4544M}, we extracted from the cosmological box all the cluster-sized halos, which were subsequently analyzed using ray-tracing techniques. Each halo was used to produce three different lens planes, obtained by projecting the cluster mass distribution along three orthogonal lines of sight. These planes of matter were used to lens a population of elliptical sources of fixed equivalent radius of $0.5$" placed on a source plane at redshift $z_{\rm s}=2$. This analysis was performed on all the halos found between $z_{\rm l}=0$ and $z_{\rm l}=2$. The strong lensing clusters were classified in three main categories, namely 1) clusters with resolved critical lines, i.e. which can produce strong lensing features such as multiple images of the same background source; 2) clusters with non-zero cross section for giant arcs, i.e. clusters which are potentially able to distort the images of background galaxies such to form arcs with length-to-width (L/W) ratios larger than 7.5; 3) what we called ``super-lenses", i.e. clusters whose lensing cross section for giant arcs is larger than $10^{-3} h^{-2}$ Mpc$^2$. This definition is based on simulations including observational noises performed with the {\tt SkyLens} code \citep{2008A&A...482..403M,2010A&A...514A..93M}, which show that, observing a cluster of galaxies with such cross section using $\sim 3$ HST orbits in the $i$-band, the expected number of giant arcs in the cluster field is $\sim 1$. Taking advantage of the large size of the cluster sample, we could characterize statistically the strong lensing cluster population in the {\sc MareNostrum Universe}, correlating the lensing strength to several cluster properties, such as their mass, shape, orientation, concentration, dynamical state, and X-ray emission. With $\sim 50,000$ strong lensing clusters found in the {\sc MareNostrum Universe}, this sample is the largest ever used for strong lensing studies.   

\section{Strong-lensing analysis of The MACS high-redshift cluster sample}
\label{sect:macs}

In a recent paper, Zitrin et al. (2010) showed the results of the
strong lensing analysis of a sample of 12 very luminous X-ray galaxy clusters at $z > 0.5$ using HST/ACS images. This is a complete sample of clusters with X-ray flux $f_X>1\times 10^{-12}$ erg s$^{-1}$ cm$^{-2}$ in the 0.1-2.4 keV band, which was defined by Ebeling et al. (2007). Later, it was targeted in several follow-up studies,
including deep X-ray, SZ and HST imaging. For example, the detection of a large-scale filament has been reported in the case of MACS J0717.5+3745 by \cite{2004ApJ...609L..49E}, for which many multiply-lensed images have been recently identified by \cite{2009ApJ...707L.102Z}, revealing this object to
be the largest known lens with an Einstein radius equivalent to $55\arcsec$ (for a source at $z\sim2.5$). In the case of MACS J1149.5+2223 \citep{2009ApJ...703L.132Z}, a background spiral galaxy at $z=1.49$ \citep{2009ApJ...707L.163S} has been shown to be multiply-lensed into several very large images, {  with a total magnification factor of $\sim 200$}. Another large multiply-lensed sub-mm source at a redshift of $z\simeq2.9$ has been identified in MACS J0454.1-0300 \citep[also referred to as MS 0451.6-0305;][]{2003PASJ...55..789T, 2004MNRAS.352..759B, 2007A&A...462..903B,2010A&A...509A..54B}, MACS J0025.4-1222 was found to be a ``bullet cluster''-like \citep{2008ApJ...687..959B}, and other MACS clusters have been recently used for an extensive arc statistics study \citep{2010MNRAS.406.1318H}. The X-ray data available for this sample (see Ebeling et al. 2007) along with the high-resolution HST/ACS imaging and additional SZ data \citep[e.g.][]{2003ApJ...583..559L} make these 12 high-redshift MACS targets particularly useful for understanding the nature of the most massive clusters.

The strong-lensing modeling of this sample, published in full in Zitrin et al. (2010) and summarized briefly here, is motivated by the successful minimalistic approach of Broadhurst et al. (2005) to lens modeling, simplified further by Zitrin et al. (2009b). This simple modeling method relies on the assumption that mass traces light so that the galaxy distribution is the starting point of the mass model, and additional flexibility between the dark matter and galaxies is allowed through the implementation of external shears. Still, the method involves only 6 free parameters, enabling easier constraints on the mass model since the number of constraints has to be equal or larger to the number of parameters in order to get a reliable fit. Two of these parameters are primarily set to reasonable values so only 4 of these parameters have to be constrained initially, which sets a very reliable starting-point using obvious systems. Recently we have further tested this assumption in Abell 1703 \citep{2010MNRAS.tmp.1308Z}, where we also applied the non-parametric technique of \cite{2006MNRAS.367.1209L} for comparison. This latter technique employs an
adaptive grid inversion method and does not make any prior assumptions of the mass distribution, yielding a very similar mass distribution to our parametric technique and hence confirming the assumption that mass generally traces light. In addition, it has been found independently that SL methods based on parametric modeling are accurate at the level of few percents at predicting the projected inner mass \citep{2010A&A...514A..93M}.

The mass distribution is therefore primarily well constrained, uncovering many multiple images which can be then iteratively incorporated into the model, by using their redshift estimation and location in the image-plane (e.g., Abell 1689, Broadhurst et al. 2005, Cl0024, Zitrin et al. 2009a). In the particular case of the 12 high-$z$ MACS clusters, in most of the clusters the multiple-images found or used in Zitrin et al. (2010a) currently lack redshift information, so that the mass profiles of most of these clusters could not be well constrained but only roughly estimated by assuming crude photometric redshifts. This however still allows an accurate determination of the critical curves for any given multiply-lensed source, as the critical curves (and the mass enclosed within them) are not dependent on the mass profile and are relatively invariant to the model parameters, enabling us to securely compare these properties to simulations as done here. In addition, note that we compare the Einstein radius for sources at $z_s\simeq2$. Due to the lensing-distance ratios, an over estimation of the source redshift by $\Delta z\sim0.5$ would only increase the projected mass and the observed Einstein radius for a source at $z_s=2$, thus resulting only in a growth of the discrepancy between observations and the $\Lambda$CDM simulations presented here. On the other hand, underestimating a source redshift would in practice decrease the observed Einstein radius and projected mass for a source at $z_s=2$ by less than 10\%, thus insignificantly influencing the results.

\section{Analysis}

\subsection{Lensing cross sections}

The efficiency of a galaxy cluster in producing arcs with a given property
can be quantified by means of its lensing cross section. This is the area on
the source plane where a source must be placed in order to be imaged as an arc
with that property.

As explained in \cite{2010arXiv1003.4544M} and in \cite{ME03.1}, the lensing cross sections are derived from the deflection angle maps, which are in turn calculated by means of ray-tracing methods \citep[see also][]{2005MNRAS.362.1301M,FE06.1}.  Bundles of light rays are traced from the observer position back to the source plane. This is populated with an adaptive grid of elliptical sources, whose spatial resolution increases towards the caustics, in order to artificially increase the number of highly magnified images. In the following analysis, a statistical weight, $w_i$, which is related to the spatial resolution of the source grid at the source position, is assigned to each source. If $a$ is the area of one pixel of the highest resolution source grid, then the area on the source plane of which the $i$-th source is representative is given by $A_i=aw_i$. The images are analyzed individually by measuring their lengths and widths using the method outlined in \cite{BA98.2} and in \cite{ME00.1}. 

We define the lensing cross section for giant arcs, $\sigma$, as
\begin{equation}
	\sigma=\sum A_i \;,
\end{equation}
where the sum is extended to all the sources producing at least one image with $L/W>7.5$.     

\subsection{Einstein rings}
While the lensing cross section is defined on the source plane, the Einstein ring is defined on the lens plane. Although the word ``ring" is appropriate only in the case of axially symmetric lenses, several authors have used it to indicate the tangential critical line of lenses with arbitrary shapes. This is the line $\vec \theta_t$ defined by the condition
\begin{equation}
\lambda_t(\vec \theta_t)=\mu_t^{-1}(\vec \theta_t)=0 \;,
\end{equation}
where $\lambda_t$ is the inverse tangential magnification, $\mu_t^{-1}$. The magnification is related to the lens convergence, $\kappa$, and shear, $\gamma$, via the equation
\begin{equation}
	\mu_t(\vec \theta)=\frac{1}{1-\kappa(\vec \theta)-\gamma(\vec \theta)} \;.
\end{equation}  
For axially symmetric lenses the following relation holds between $\kappa$ and $\gamma$:
\begin{equation}
	\gamma(\theta) = \overline\kappa(\theta)-\kappa(\theta) \;,
	\label{eq:gammacirc}
\end{equation}
where $\overline\kappa(\theta)$ indicates the mean convergence within a circle of radius $\theta$. Using the above formulas, the Einstein ring of an axially symmetric lens is defined as the distance $\theta_E$ from the lens center where 
\begin{equation}
	1-\overline\kappa(\theta_E)=0 \;,
	\label{eq:thetae}
\end{equation}
i.e. as the radius of a circle enclosing a mean convergence of 1. We remind that the convergence is related to the lens surface density $\Sigma$ and to the critical surface density for lensing $\Sigma_{\rm cr}$ via the equation
\begin{equation}
	\kappa(\vec \theta)=\frac{\Sigma(\vec \theta)}{\Sigma_{\rm cr}} \;.
\end{equation}
Thus, the mean surface density within the Einstein ring of an axially symmetric lens is 
\begin{equation}
	\overline\Sigma_E=\Sigma_{\rm cr}=\frac{c^2}{4 \pi G}\frac{D_{\rm s}}{D_{\rm ls}D_{\rm l}} \;,
\end{equation}
where the $D_{\rm s}$, $D_{\rm l}$, and $D_{\rm ls}$ denote the angular diameter distances between the observer and the source plane, between the observer and the lens plane, and between the lens and the source planes, respectively.

\begin{figure}[lt!]
\begin{center}
  \includegraphics[width=\hsize]{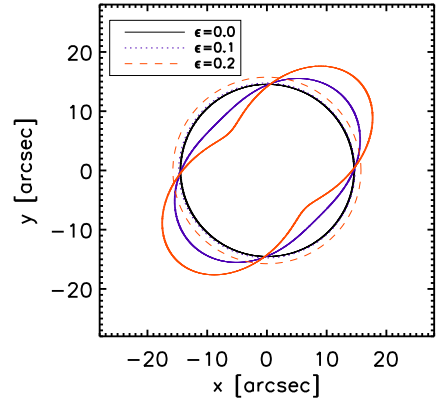}
\end{center}
\caption{The median Einstein rings of a lens with increasingly larger ellipticity of the lensing potential. Black, blue, and red solid lines show the tangential critical lines of lenses with ellipticity 0, 0.1, and 0.2, respectively. The corresponding median Einstein rings are given by the solid, dotted, and dashed lines of the same colors.}
\label{fig:rmedell}
\end{figure}

\begin{figure}[lt!]
\begin{center}
  \includegraphics[width=\hsize]{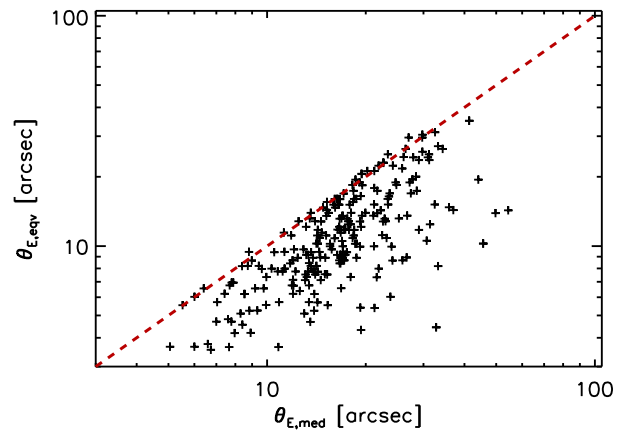}
\end{center}
\caption{The comparison between {\rm equivalent} and {\em median} Einstein ring sizes for a sample of clusters with mass $M>5\times 10^{14} h^{-1}\,M_\odot$ at $z>0.5$ extracted from the {\sc MareNostrum Universe}. Each cross represents a cluster projection. The dashed red line indicate the bisector of the $\theta_{\rm E,med}-\theta_{\rm E,eqv}$ plane.}
\label{fig:recomp}
\end{figure}

This is strictly applicable only to axially symmetric lenses, but this definition of Einstein ring has been exported to arbitrary lenses by several authors \citep[see e.g][]{2010arXiv1002.0521Z,2010MNRAS.404..325R}, which define the {\em equivalent} Einstein ring size, $\theta_{\rm E,eqv}$, via Eq.~\ref{eq:thetae}. In this paper, we follow a different approach. We define a {\em median} Einstein ring size, $\theta_{\rm E,med}$, which is  the median distance of the tangential critical points from the cluster center\footnote{  When dealing with mass distributions characterized by asymmetries and substructures, the cluster center may not be easily defined. In this work, the cluster center is determined by smoothing the projected cluster mass distribution with a gaussian kernel with a FWHM of $30$ kpc. This is meant to erase the local peaks related to the presence of galaxy-scale halos. The center of the cluster is then defined as the location of the maximum of the smoothed projected mass distribution. This is only one of the possible choices for the cluster center. Of course, the same definition holds for simulated and real clusters.} :
\begin{equation}
\theta_{\rm E,med}={\rm median}(\vec\theta_t)\;.
\label{eq:remed}
\end{equation}
 This definition better captures the important effect of shear caused by the cluster substructures, whose effect is that of elongating the tangential critical lines along preferred directions, pushing the critical points to distances where $\kappa$ is well below unity \citep[see also][]{BA95.1}. For example, an axially symmetric lens embedded in an external shear has a larger $\theta_{\rm E,med}$ compared to an isolated axially symmetric lens. The same argument applies to lenses whose lensing potential is elliptical rather than spherical. This is shown in Fig.~\ref{fig:rmedell}, where we display the tangential critical lines of a lens with NFW density profile whose iso-potential contours have ellipticities $\epsilon=\frac{a^2-b^2}{a^2+b^2}$ equal to 0. 0.1, and 0.2. The solid, dotted, and dashed lines show the respective {\em median} Einstein rings, whose radius is derived from Eq.~\ref{eq:remed}.

\begin{figure*}[lt!]
\begin{center}
  \includegraphics[width=0.49\hsize]{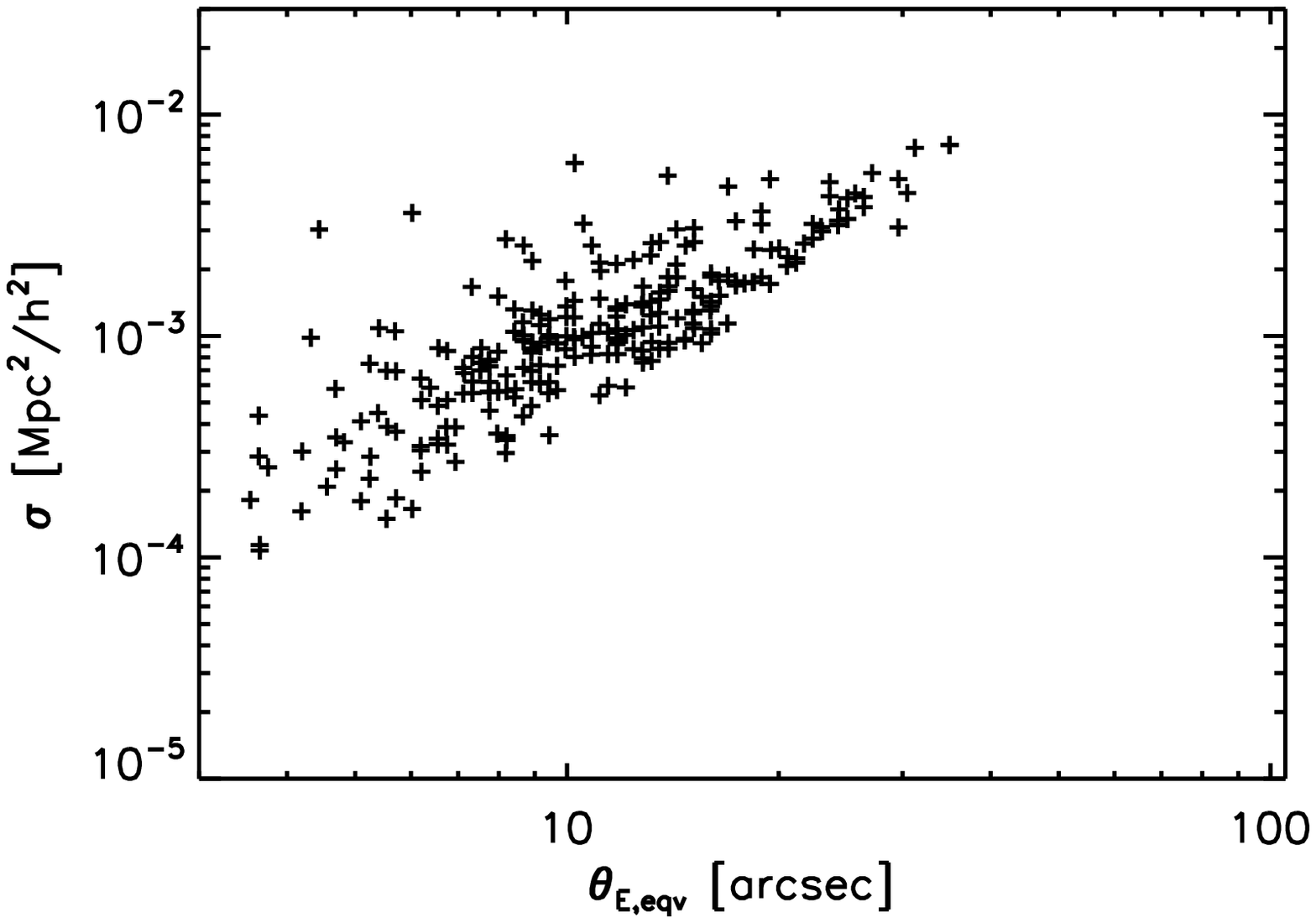}
  \includegraphics[width=0.49\hsize]{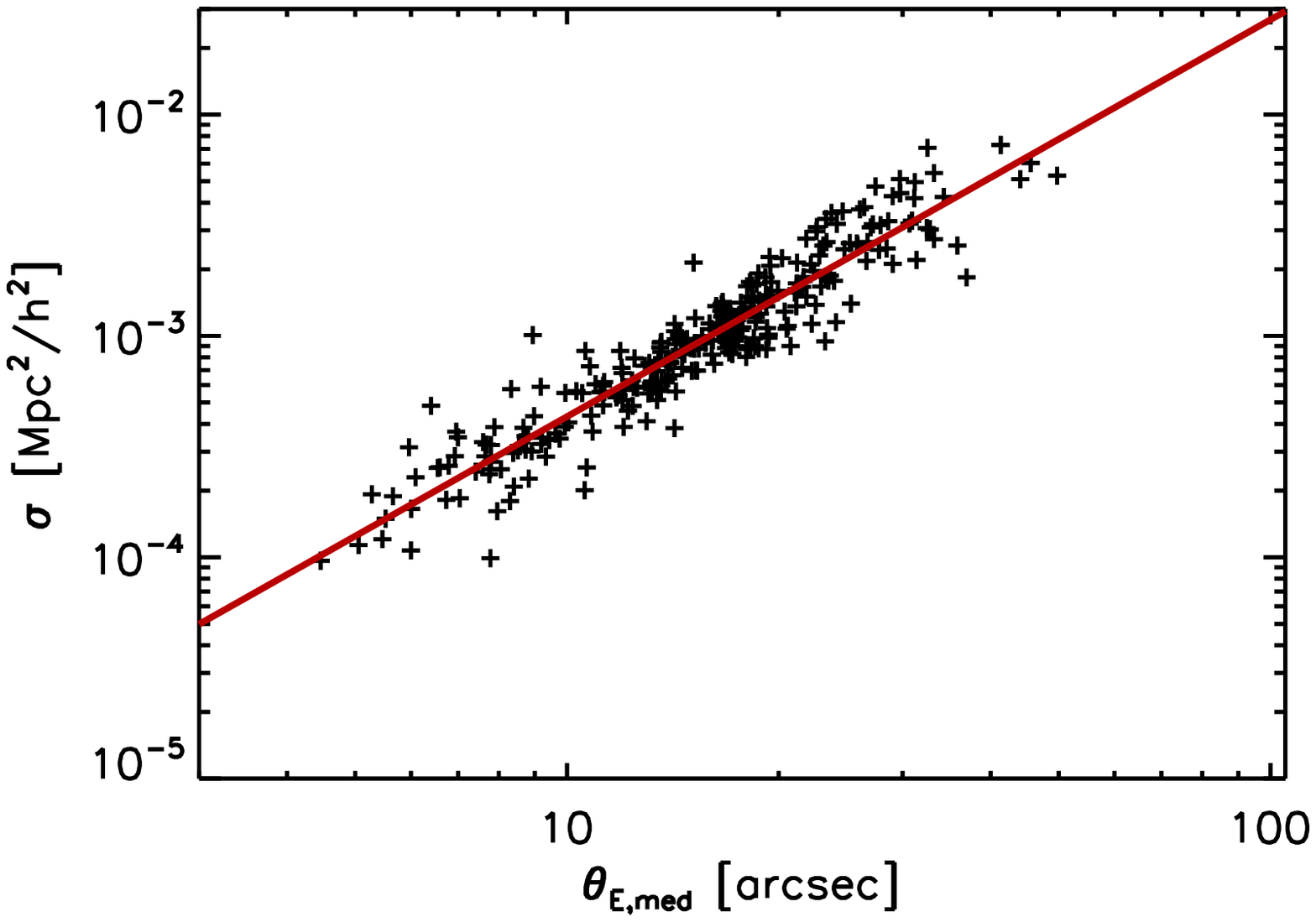}
\end{center}
\caption{Lensing cross section for giant arcs vs Einstein ring size for a sample of clusters with mass $M>5\times 10^{14} h^{-1}\,M_\odot$ at $z>0.5$ extracted from the {\sc MareNostrum Universe}. The two panels refer to the two different definitions of the Einstein ring radius. On the left, we use the {\em equivalent} Einstein radius, while on the right we use the {\em median} Einstein radius. The red solid line in the right panel indicates the best linear fit relation between $\log(\sigma)$ and $\log(\theta_{\rm E,med})$.}
\label{fig:recs}
\end{figure*}
In Fig.~\ref{fig:recomp}, we show a comparison between the two definitions of Einstein ring size, when they are applied to a sample of massive clusters ($M>5\times 10^{14} h^{-1}\,M_\odot$) at redshift $z>0.5$ taken from the {\sc MareNostrum Universe}. Each cross represents a cluster projection in the $\theta_{\rm E,med}-\theta_{\rm E,eqv}$ plane. Clearly, almost all the points lay in the bottom-right part of the diagram, below the bisector shown by the dashed red line. Thus, $\theta_{\rm E,med}\gtrsim \theta_{\rm E,eqv}$ for the vast majority of the clusters. This indicates that these systems are typically non-axially symmetric and contain many substructures that enhance their shear fields. 
\cite{ME07.1} showed that asymmetries and substructures contribute significantly to the lensing cross section for giant arcs. Therefore, we expect that $\theta_{\rm E,med}$ correlates much better with $\sigma$ than $\theta_{\rm E,eqv}$ does. This is shown in Fig.~\ref{fig:recs}. In the left and in the right panels we plot the lensing cross section vs the equivalent and the median Einstein ring size for the same clusters used in Fig.~\ref{fig:recomp}.     
The figure shows that, when the {\em median} Einstein radius is used, all the data points lay very close to a line in the $\log(\sigma)-\log(\theta_{\rm E,med})$ plane, whose equation is
\begin{equation}
\log(\sigma)=(1.79\pm0.04)\log(\theta_{\rm E,med})-(5.16 \pm 0.05)\;.
\label{eq:bf}
\end{equation}
The Pearson correlation coefficient is $r=0.94$, confirming that the correlation between the two plotted quantities is very strong.

As shown in the left panel, using the {\em equivalent} Einstein radius, the scatter is substantially larger and the correlation in much worse. The Pearson coefficient is $r=0.75$ in this case. For this reason, in the following analysis we prefer to use the {\em median}, rather than the {\em equivalent} Einstein radius. 

The tight correlation existing between $\sigma$ and $\theta_{\rm E,med}$ highlights the strong connection between the arc statistics and the Einstein ring problems. If the Einstein ring sizes were too large for the $\Lambda$CDM model, then we would also observe an excess of giant arcs compared to the expectations. 

\subsection{Analysis of the MACS sample}
As we pointed out in Sect.~\ref{sect:intro}, our purpose is to perform a consistent comparison between real and simulated galaxy clusters. For this reason, we use the deflection angle maps obtained for the MACS cluster reconstructions of \cite{2010arXiv1002.0521Z} and we use them to perform lensing simulations with the same methods used to analyze the clusters in the {\sc MareNostrum Universe}. For each cluster, we measure the lensing cross section and the Einstein radius. Table~\ref{tab:ersigma} summarizes the results found for the 12 MACS clusters used in this paper. For each cluster in the sample, we also list the redshift in the second column. Note that all but two clusters have lensing cross sections for giant arcs $\sigma>10^{-3}\,h^{-2}$Mpc$^2$. On the basis of the classification proposed in \cite{2010arXiv1003.4544M}, these clusters could be classified as super-lenses. 

\begin{table}[htdp]
\caption{The lensing cross sections, the median and the equivalent Einstein radii of the 12 MACS clusters used in this work. The values reported for $\theta_{\rm E,eqv}$ are taken from Table 2 of \cite{2010arXiv1002.0521Z}.}
\begin{center}
\begin{tabular}{ccccc}
\hline
\hline
MACS & $z$ & $\sigma$ & $\theta_{\rm E,med}$ & $\theta_{\rm E,eqv}$ \\ 
        & & [$10^{-3}\,h^{-2}$Mpc$^2$] & [arcsec] & [arcsec] \\
\hline
J0018.5+1626 & 0.545 & 3.27 & 31.5 & 24\\
J0025.4-1222 & 0.584 & 15.27 & 55.8 & 30\\ 
J0257.1-2325 & 0.505 & 7.89 & 54.2 & 39\\ 
J0454.1-0300 & 0.538 & 0.98 & 13.9 & 13\\
J0647.7+7015 & 0.591 & 3.07 & 27.5 & 28\\
J0717.5+3745 & 0.546 & 14.2 & 71.1 & 55\\
J0744.8+3927 & 0.698 & 2.30 & 32.2 & 31\\
J0911.2+1746 & 0.505 & 0.25 & 10.5 & 11\\
J1149.5+2223 & 0.544 & 2.69 & 25.0 & 27\\
J1423.8+2404 & 0.543 & 1.64 & 21.5 & 20\\
J2129.4-0741 & 0.589 & 8.48 & 46.0 & 37\\
J2214.9-1359 & 0.503 & 2.22 & 26.0 & 23\\
\hline
\hline
\end{tabular}
\end{center}
\label{tab:ersigma}
\end{table}%

\begin{figure}[lt!]
\begin{center}
  \includegraphics[width=\hsize]{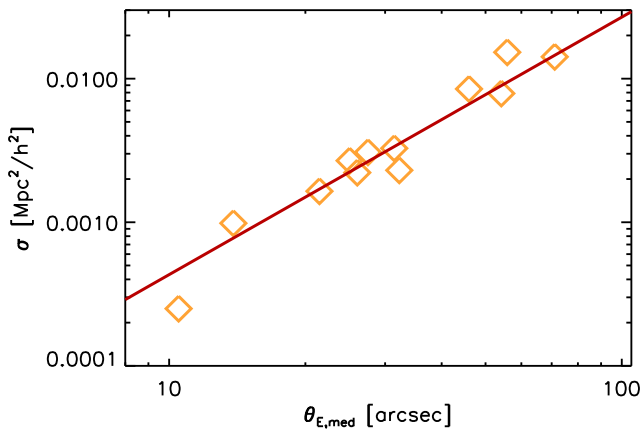}
\end{center}
\caption{The $\sigma-\theta_{\rm E,med}$ relation for the 12 MACS clusters used in this work. Each diamond indicates a cluster, while the red line is the best-fit relation found for the clusters in the {\sc MareNostrum Universe} and given in Eq.~\ref{eq:bf}.} 
\label{fig:csthetamacs}
\end{figure}

It is very interesting to note that the $\theta_{\rm E,med}-\sigma$ relation found for the MACS clusters is consistent with that measured in the {\sc MareNostrum Universe}. This is shown in Fig.~\ref{fig:csthetamacs}. Each diamond indicates the position of a MACS cluster in the $\theta_{\rm E,med}-\sigma$ plane. The red solid line shows the best fit relation given in Eq.~\ref{eq:bf}, derived from the clusters in the {\sc MareNostrum Universe}. The MACS clusters nicely follow the same relation found in the simulations. 
  
\section{Comparison between simulations and observations}
We now proceed at comparing statistically the strong lensing properties of the MACS clusters to those of the clusters in the {\sc MareNostrum Universe}. 

\subsection{Correction of the X-ray luminosities}
The MACS sample is an X-ray flux selected sample. The selection method is discussed in \cite{2007ApJ...661L..33E} and summarized later in this paper. 
As explained in \cite{2010arXiv1003.4544M} we have measured the X-ray emission of all clusters in the {\sc MareNostrum Universe}, which, in principle, would allow us to apply the same selection function used for the MACS sample to the simulated clusters. However, the {\sc MareNostrum Universe} is non-radiative simulation and it is well known that the X-ray properties of galaxy clusters can be reproduced only by means of a more sophisticated description of the gas physics. In particular, the X-ray luminosity of low mass clusters is known to be over-predicted in adiabatic simulations, resulting in a much shallower $M-L_X$ relation than observed. A discussion about how the $M-L_X$ relation changes depending on the different gas physics can be found in \cite{2010arXiv1002.4539S}. The authors of this paper show that only including cooling and some mechanism, like pre-heating or AGN feedback, to heat the gas and prevent it from reaching high central densities, can the simulation match the slope of the observed $M-L_X$ relation at low redshift, as derived from the REXCESS data \citep{2009A&A...498..361P}. They also provide analytic formulas for describing the redshift evolution of the X-ray scaling relations in different kind of simulations. In particular, the $M-L_X$ relation is parametrized as follows:
\begin{equation}
E(z)^{-7/3}L_{X}=C(z)\left(\frac{M}{M_0}\right)^\alpha \;,
\label{eq:mlrel}
\end{equation}  
where 
\begin{equation}
E(z)=\sqrt{\Omega_{\rm m,0}(1+z)^3+\Omega_{\rm \Lambda,0}}
\end{equation}
for a spatially flat $\Lambda$CDM cosmological model, and 
\begin{equation}
C(z)=C_0(1+z)^\beta \;.
\end{equation}  
The parameters $C_0$, $\alpha$, and $\beta$ are estimated by fitting the scaling relations of the simulated clusters. The best fit values are reported in Table~2 and 3 of \cite{2010arXiv1002.4539S} for different gas physics implemented in the simulations. The mass $M_0$ is $5\times10^{14}h^{-1}\,M_\odot$. 

Using an high redshift sample of X-ray clusters from \cite{2008ApJS..174..117M}, the authors find that the scaling relations derived by including an AGN feedback model (FO run) evolve broadly consistently with the observational data. From  Eq.~\ref{eq:mlrel} and the best fit values found by \cite{2010arXiv1002.4539S} for their adiabatic (GO) and FO runs, we derive the correction we should apply to the X-ray luminosities of our simulated clusters in order to facilitate a comparison to the observations. This correction is estimated as 
\begin{equation}
f_{\rm corr}=\frac{L_{X,{\rm corr}}}{L_X}=0.316(1+z)^{0.991}\left(\frac{M}{M_0}\right)^{0.574} \;.
\end{equation}

\subsection{Simulating MACS-like surveys}
We use the {\sc MareNostrum Universe} as a reference to build up mock catalogs of MACS-like clusters. The simulated clusters are distributed on the sky within a spherical shell between $z=0.5$ and $z=0.7$. This volume is subdivided into seven sub-shells, which are populated with objects taken from the  snapshots at $z=0.5,0.53,0.56,0.59,0.63,0.66$ and $0.68$ of the simulation. When necessary, we replicate the clusters in the cosmological box such to reproduce the expected number of halos in each shell. The number of replicates is determined by the ratio between the shell volume and the simulation volume. For the lensing analysis, each cluster is projected along three different lines of sight. Every time we replicate a cluster, we randomly choose the line of sight along which it is observed. 

The sample of MACS clusters used in this work was constructed by using the following selection criteria:
\begin{itemize}
\item the X-ray flux in the band [0.1-2.4] keV is $f_{X}>1\times10^{-12}$ erg s$^{-1}$ cm$^{-2}$;
\item the clusters are observable from Mauna Kea: $|b|\ge 20^\circ$, $-40^\circ \le \delta \le 80^\circ$;
\item the redshift is $z>0.5$ (the most distant cluster is at $z=0.698$, as shown in Table~\ref{tab:ersigma}).
\end{itemize}
Applying the cuts to our mock cluster catalogs, we finally define a MACS-like sample for our comparison. We repeat the procedure to generate the cluster catalogs $10$ times to partially account for the cosmic variance.      

\subsection{Distributions of lensing cross sections}

\begin{figure}[lt!]
\begin{center}
  \includegraphics[width=\hsize]{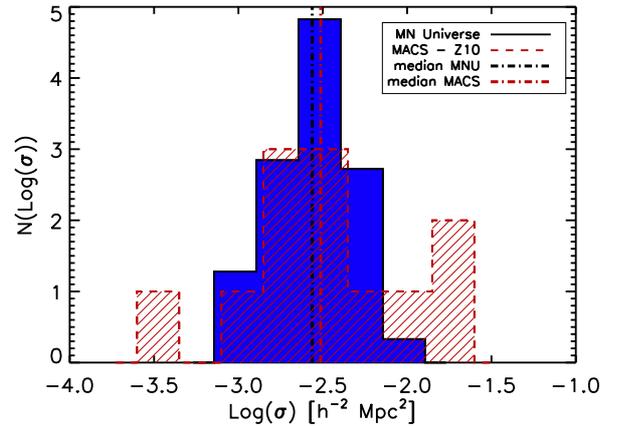}
\end{center}
\caption{The distributions of the strong lensing cross sections. The blue filled histogram shows the results for the simulated MACS sample constructed with clusters taken from the {\sc MareNostrum Universe}. The red shaded histogram shows the same distribution but for the observed MACS sample. The vertical dot-dashed lines indicate the medians of the two distributions.} 
\label{fig:cshist}
\end{figure}

A comparison between the distributions of the lensing cross sections for giant arcs in the simulated and in the observed MACS samples is shown in Fig.~\ref{fig:cshist}. The histograms have been normalized to the number of clusters in the observed MACS sample (12).
A two-sided Kolmogorov-Smirnov test returns a probability of $\sim 12\%$ that the observed and simulated data are drawn from the same distribution.
The most remarkable difference between the two samples is the excess of clusters with large lensing cross section among the MACS clusters. Three clusters, namely  MACSJ0717.5+3745, MACSJ0025.4-1222, and MACSJ2129.4-0741, have lensing cross sections larger than any cluster in the {\sc MareNostrum Universe}. The medians of the two distributions are $\sigma_{\rm med,MACS}=3.07\times10^{-3}\,h^{-2}$Mpc$^2$ and $\sigma_{\rm med,MNU}=2.7\times10^{-3}\,h^{-2}$Mpc$^2$ for the observed and for the simulated MACS samples, respectively. The lensing cross section of cluster MACSJ0911.2+1746 is a factor of $\sim 2$ smaller than the smallest cross section among the simulated clusters. As shown in Fig.~\ref{fig:csthetamacs}, this cluster lays below the predicted $\log{\sigma}-\log{\theta_{\rm E,med}}$ scaling relation, i.e. the lensing cross section is small given the size of the Einstein ring.  

The distributions can be used to estimate the differences between the expected number of giant arcs produced by the two cluster samples (when they lens sources at redshift $z_{\rm s}=2$). The number of arcs given by
\begin{equation}
N_{\rm arcs}=n_{\rm s} \sum_{i=1}^{n_{\rm clus}} \sigma_i \;,
\label{eq:narcs}
\end{equation}
where $n_{\rm s}$ is the number density of sources in the background of the clusters, $\sigma_i$ is the lensing cross section of the $i-th$ cluster in the sample, and the sum is extended to all the clusters in the sample. Using Eq.~\ref{eq:narcs}, we estimate that about a factor $\sim 2$ less arcs are expected from the {\sc MareNostrum} clusters compared to what expected from the MACS clusters. This is far from the order-of-magnitude difference found by \cite{BA98.2} between clusters simulated in the framework of the $\Lambda$CDM cosmological model and observations, reducing substantially the size of the arc-statistics problem. This is not surprising since our simulations include the effects of mergers \citep{TO04.1,FE06.1}, which could not be properly taken into account by \cite{BA98.2} due to the limited number of clusters in their sample and to coarse time resolution in their simulations. Analyzing the {\sc MareNostrum Universe}, we showed in \cite{2010arXiv1007.1551F} that un-relaxed clusters contribute to $\sim 70\%$ of the optical depth
for clusters at $z>0.5$.

\subsection{Distributions of Einstein ring sizes}
\begin{figure}[lt!]
\begin{center}
  \includegraphics[width=\hsize]{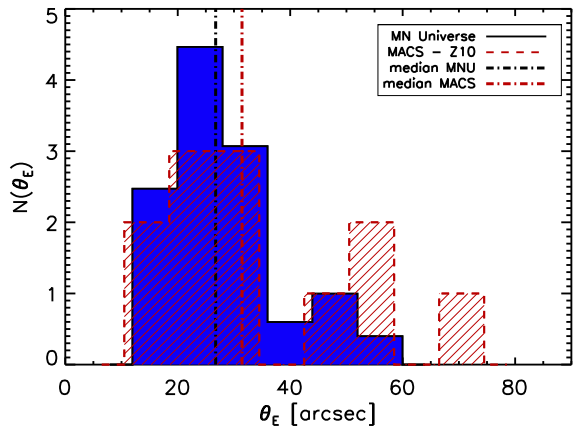}
\end{center}
\caption{The distributions of the Einstein ring sizes. The blue filled histogram shows the results for the simulated MACS sample constructed with clusters taken from the {\sc MareNostrum Universe}. The red shaded histogram shows the same distribution but for the observed MACS sample. The vertical dot-dashed lines indicate the medians of the two distributions.} 
\label{fig:rehist}
\end{figure}

In Fig.~\ref{fig:rehist} we compare the distributions of the Einstein ring sizes for the simulated and the observed MACS samples. As found for the lensing cross sections, the distribution derived for the observed cluster sample is also characterized by an excess of clusters towards the large values of $\theta_{\rm E}$. The same clusters having larger cross sections compared to any simulated cluster, also have Einstein radii exceeding the maximum value found in the simulations. The Kolmogorov-Smirnov test returns a probability of $\sim 30\%$ that the two datasets are drawn from the same statistical distribution.
The Einstein ring medians  of the observed and of the simulated samples differ by $\sim 25\%$, while \cite{2010arXiv1002.0521Z} report a difference of $\sim 40\%$ between the observed and the theoretical distributions of Einstein radii. However, their theoretical estimates are based on analytic models of galaxy clusters, which are described by means of axially symmetric lenses with NFW density profiles. \cite{ME03.1} showed that such simple lens models under-estimate the lensing cross sections for giant arcs of numerically simulated clusters, mainly because they miss several important features like asymmetries and substructures, which enhance significantly the lensing efficiency of galaxy clusters \citep[see also][]{ME07.1}. Given the strong correlation existing between lensing cross section and Einstein radius, shown in Fig.~\ref{fig:recs}, it is not surprising that our numerically simulated galaxy clusters have larger Einstein radii. Thus, as the arc statistics problem, also the Einstein ring problem results to be alleviated on the basis of our results, although three clusters still do not have a counterpart in our $\Lambda$CDM simulation.

\subsection{Expected concentration bias}
\cite{2010arXiv1003.4544M} showed that, due to the cluster triaxiality and to the orientation bias that affects the strong lensing cluster population, we should expect to measure substantially higher concentrations than expected in clusters showing many strong lensing features \citep[see also][]{2005ApJ...632..841O,2009ApJ...699.1038O}. Indeed, the amplitude of this bias is a growing function of the lensing cross section. It is worth (and easy) to estimate the bias which is expected to affect a MACS-like sample of clusters. For each of the clusters in the simulated sample we measure the concentration inferred from both the projected and the three-dimensional mass distributions, $c_{\rm 2D}$ and $c_{\rm 3D}$. These are measured by the surface mass density and the three-dimensional mass density with NFW models. The concentration bias is quantified by means of the ratio between the two-dimensional and the three-dimensional concentrations.     

\begin{figure}[lt!]
\begin{center}
  \includegraphics[width=\hsize]{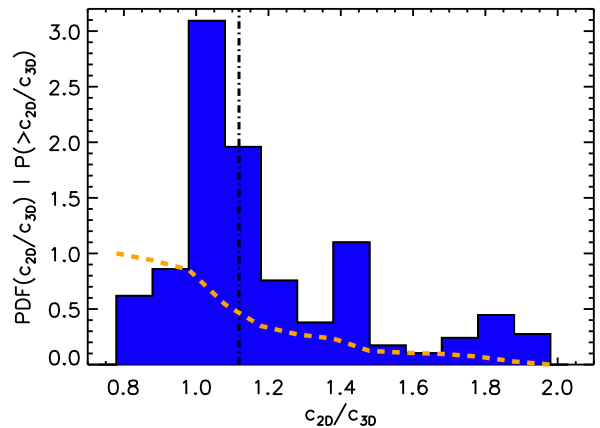}
\end{center}
\caption{The probability density function of the ratios between 2D and 3D concentrations for clusters with the same properties of those in the MACS sample. The vertical dot-dashed line indicates the median of the distribution. The dashed line shows the cumulative distribution.} 
\label{fig:concb}
\end{figure}

The distribution of $c_{\rm 2D}/c_{\rm 3D}$ derived from the simulated MACS sample is shown in Fig.~\ref{fig:concb}. As expected, a median bias of the order of $\sim 11\%$ is found. However, the distribution is skewed towards the high values. On the basis of our simulations we expect that the concentration may be biased by up to $100\%$ for some of the clusters. The dashed line shows the cumulative probability distribution. For about $\sim 20\%$ of the clusters in the simulated sample we measure a 2D concentration which is $>40\%$ higher than the 3D concentration. 

\begin{figure}[lt!]
\begin{center}
  \includegraphics[width=\hsize]{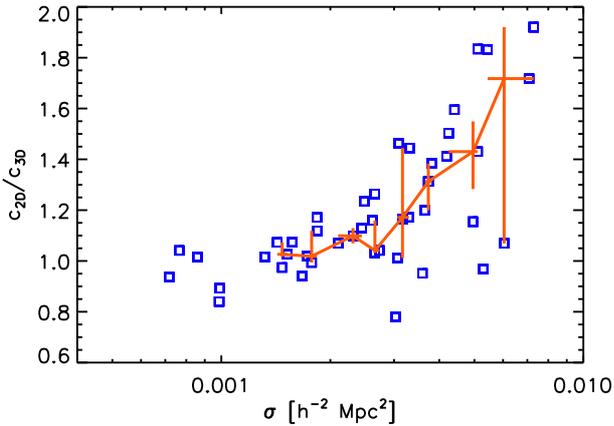}
\end{center}
\caption{The concentration bias vs the strong lensing cross section. Each square represents a simulated cluster from a mock MACS catalog. The solid line is the median relation, while the error-bars indicate the inter-quartile ranges in each bin.} 
\label{fig:ccs}
\end{figure}

In Fig.~\ref{fig:ccs}, we show how the concentration bias depends on the lensing cross section. While the bias is very small, of the order of few percent, for clusters with $\sigma<10^{-3}\,h^{-2}$Mpc$^2$, it becomes increasingly higher for the clusters with larger cross sections. This is a very important result for interpreting the high concentrations recently measured in clusters like A1689 or CL0024 \citep{2008ApJ...685L...9B,2009MNRAS.396.1985Z}. 

\section{Discussion and conclusions}
In this paper, we have used a large set of numerically simulated clusters, for which both the X-ray and the lensing properties were previously investigated, and we have compared their Einstein ring sizes {  and } strong lensing cross sections to those of clusters in an X-ray selected sample at redshift $z>0.5$ (MACS). {  We have also used the numerical simulations to estimate the expected bias in the lensing-inferred concentrations of the same MACS sample.} Our aim was to investigate whether some previously claimed discrepancies between the strong lensing properties of real and analytically modeled clusters in the framework of $\Lambda$CDM cosmology were confirmed by adopting more sophisticated and realistic cluster mass distributions. These discrepancies between theory and observations regard 1) the large observed number of gravitational arcs behind the cores of galaxy clusters, 2) the large size of the Einstein rings reconstructed through lensing-based observations, and 3) the large mass concentrations determined by fitting the two-dimensional mass distribution of strong lensing clusters. Our main results can be summarized as follows:
\begin{itemize}
\item the comparison between the numerical and the observational cluster sample shows that some real clusters have too large lensing cross sections and Einstein rings compared to expectations in a $\Lambda$CDM cosmological model. However, the discrepancy between observations and simulations is now significantly reduced compared to previous studies based on analytical estimates. Using N-body and hydrodynamical simulations allows to better capture several important properties of strong lensing clusters, such as their triaxiality, asymmetries, concentration scatter, dynamical activity, all of which were proven to boost the lensing efficiency;
\item using numerically simulated clusters we can predict the lensing concentration bias expected for the MACS sample. Such an estimate is based on the assumption that the numerical models are representative of the real cluster population. As shown by our results on the lensing cross sections and on the size of the Einstein rings, this may not be the case. Measuring a larger-than-expected concentration bias in real clusters may be a further indication that some substantial difference exists between the inner structure of observed and simulated clusters.
\end{itemize}

We believe that these results are important because they  suggest that our simplified models of the physical processes affecting the matter distribution in the cores of galaxy clusters must be improved. First of all this improvement concerns the description of gas-dynamical processes in clusters. On the other hand, our results may also indicate that some fundamental assumption in the $\Lambda$CDM model is incorrect and leads to wrong predictions of the matter distribution on scales probed by galaxy clusters. In this case, alternative cosmologies, involving dynamical dark-energy, modified gravity, or primordial non-gaussianity may better explain the differences between the strong lensing properties of simulated and observed galaxy clusters. 

However, a few aspects of our work deserve some further discussion and deeper investigation.
The simulations used in this study are performed in the framework of a WMAP-1 normalized cosmology. The value of $\sigma_8$ used in this study is larger than measured in the WMAP-7 data release. If the WMAP-7 normalization was adopted, the differences between simulations and observations would be significantly amplified. \cite{FE08.1} show that, decreasing $\sigma_8$ from 0.9 to 0.8, the average lensing optical depth drops by at least a factor of three. The cluster efficiency for strong lensing would be strongly dimmed at $z>0.5$, as a result of the delayed structure formation caused by a lower normalization of the CDM power-spectrum.

Our simulations suffer of two important limitations. The first is certainly given by the size of the cosmological box. Due to the relatively small box size, we may be under-predicting the number of very high-mass clusters in the sample. Moreover, we may be missing some systems that are particularly dynamically active. Consequently, our numerical sample may miss some of the most powerful lenses. We will investigate this issue by using a sample of massive clusters extracted from a larger cosmological box.  The second limitation is due to the simple description of the gas physics in the simulation, where no radiative and feedback processes are considered. On the basis of the recent results of \cite{2010MNRAS.tmp..671M}, we are confident that this does not affect significantly the lensing properties of the clusters. It has been shown that the energy feedback from AGNs and supernovae counter-act and compensate the effects of cooling on the strong lensing cross sections. However, as we explained earlier, the X-ray scaling relations are strongly affected by such physical processes. We introduced a correction to the X-ray fluxes based on the comparison between simulations and observations for accounting for this poor description of the gas physics. In absence of more sophisticated cluster models, this was the only viable approach, which is however based on several approximations.

As mentioned in Sect.~\ref{sect:macs}, most of the multiple image systems detected behind the MACS clusters do not have spectroscopically confirmed redshifts. This lack of information makes these cluster mass reconstructions uncertain, although we believe that our conclusions would be not affected dramatically {  (see the discussion in the Appendix~\ref{app:clmod})}. This situation will be soon improved thanks to the upcoming Hubble Multi-Cycle-Treasury-Program {\tt CLASH}\footnote{{\tt http://http://www.stsci.edu/$\sim$postman/CLASH/}}, which will dedicate 524 new HST orbits to observe 25 galaxy clusters in 16 different bands, spanning the near-UV to near-IR (P.I. Postman). These observations, combined with already existing data for some of the targets,  are expected to deliver $\sim 40$ new multiple image systems per cluster with photometric redshift determined with an accuracy $\Delta z \sim 0.02(1+z)$. Half of the clusters in the MACS sample used in this paper are also in the CLASH target list. As shown by \cite{2010A&A...514A..93M}, these observations will allow to constrain the Einstein ring sizes and the projected mass distributions in the inner regions of clusters with accuracies at the percent level. We look forward to improving our comparison between observations and simulations using this new dataset as well as improved cosmological simulations.

While our paper was in the process of being refereed, \cite{2011arXiv1101.4653H} posted a paper were a comparison between the lensed ac statistics by simulated halos taken from the Millennium simulation \citep{2005Natur.435..629S} and by a sample of X-ray selected clusters is shown. Their results for $z>0.5$ agree with ours, being the simulated arc production efficiency lower by a factor of 3 than observed in the MACS cluster sample. Instead, at lower redshift ($z \sim 0.4$), they find a very good agreement between the observed and the simulated arc statistics, in terms of the mean number of arcs per cluster, the distribution of number of arcs per clusters, and the angular separation distribution. At even lower redshift, ($z\sim 0.2$) they again find an excess of arcs in the observations compared to simulations.  Clearly, this emphasises that we need much larger samples before we arrive at a firm conclusion.  

\acknowledgements{MM thanks Elena Rasia and Pasquale Mazzotta for their useful comments on this work. We are grateful to M. Limousin for having carefully revised our paper and for having provided us the models of some some of the MACS clusters. We acknowledge financial contributions from contracts ASI-INAF I/023/05/0, ASI-INAF I/088/06/0, and PRIN INAF 2009. The {\sc MareNostrum Universe} simulation was done at the BSC-CNS
(Spain) and analyzed at NIC J\"ulich (Germany). SG acknowledges the
support of the European Science Foundation through the ASTROSIM Exchange Visits Program. GY acknowledges support of  MICINN  (Spain) through research grants FPA2009-08958 and  AYA2009-13875-C03-02.}
\bibliography{./master}
\bibliographystyle{aa}

{ 
\appendix
\section{How stable are the MACS lens models?}
\label{app:clmod}
As we pointed out earlier, the major source of uncertainty in the mass models of the MACS clusters used in this study is the lack of spectroscopic redshifts available for most of the multiple image systems. Indeed, for several of these systems the redshifts are estimated photometrically. Other groups who performed a strong lensing analyses on some of the clusters used in this study did not agree on the identification of several strong lensing systems. In order  to have an idea of how stable the models are with respect to the assumptions made, we compare here some of the models of \cite{2010arXiv1002.0521Z} (Z models in the following) with alternative models kindly provided by Marceu Limousin. It is important to note that 1) these models are obtained with a completely different mass reconstruction code, the public software {\tt lenstool}\footnote{see \tt{http://www.oamp.fr/cosmology/lenstool/}}, which follows a completely different approach compared to    
\cite{2010arXiv1002.0521Z}, and 2) in some cases the associations and the number of multiple image systems differ significantly.

The {\tt lenstool} models (L models hereafter) were provided for four of the MACS clusters under investigation. These are the clusters MACSJ0717.5+3745 (Limousin et al., in prep.), MACSJ1149.5+2223 \citep[see details in ][]{2009ApJ...707L.163S}, MACSJ0454.1-0300 \citep{2007A&A...462..903B,2010A&A...509A..54B}, and MACSJ1423.8+2404 \citep{2010MNRAS.405..777L}. We use these models fully consistently to the analysis made on the Z reconstructions, in order to derive measurements of the Einstein rings and of the lensing cross sections. 

The L model of MACSJ0717 is based on a set of 15 multiple image systems, the vast majority of which agree with the systems identified and used by \cite{2010arXiv1002.0521Z}. Recently Limousin and collaborators gathered spectroscopic redshifts for two of these systems, which have been used to calibrate their model. The resulting reconstruction has and Einstein ring $\theta_{\rm med}=61''$ and a lensing cross section $\sigma=1.24\times10^{2}h^{-2}$Mpc$^{-2}$. These values are only $\sim 15\%$ smaller than those of the corresponding Z model.

In MACSJ1149, the most spectacular lensed system is a spiral galaxy
whose total magnification is estimated to be $\sim 200$. According to \cite{2009ApJ...703L.132Z} and \cite{2010arXiv1002.0521Z} (Z model), this galaxy
is lensed into five images, while \cite{2009ApJ...707L.163S} refer to the fifth image as part of the fourth image. This of course is a matter of interpretation. The calibration of the Z model is based on the assumption that the redshift of the above mentioned spiral galaxy is $\sim 1.5$, which was then confirmed spectroscopically by \cite{2009ApJ...707L.163S}.
Additionally, the Z lens model itself has allowed to identify six additional multiple-image candidate systems, that are not used in the construction of the L model nor are plausible in this model. The mass reconstructions derived from the constraints used by the two groups correspondingly differ.
In particular, the Z model has a shallower projected
density profile than the L model. Despite this, the two
models result to be almost identical in terms of both Einstein ring
and lensing cross section, with differences of order of only a few
percent.

Using the L model of MACSJ1423 we find an Einstein ring and a cross section that are respectively $\sim 30\%$ and $\sim 25\%$ smaller than those derived from the Z model. Strangely, for this cluster spectroscopic redshifts exist for two sets of multiple images which are used to build both the L and the Z models, thus such a mismatch between the models was quite unexpected.
However, by visually inspecting the critical lines of the two reconstructions, we notice that the Z model predicts a northward extension of the tangential critical line, surrounding a bright elliptical galaxy \citep[see Fig. 19 of ][]{2010arXiv1002.0521Z}. Such extension is not present in the L model and contributes to increasing both $\theta_{\rm E}$ and $\sigma$. It may be an artifact or not, depending on the still unknown weight of this galaxy, as there are no lensing constraints in that region of the lens plane which probe this portion of the critical curve. 

MACSJ0454 shows the largest discrepancy between L and Z models. In this case, the Einstein ring size and the lensing cross section derived from the L model are extremely larger than those derived from the Z model (a factor of two and a factor of five, respectively). We do not fully understand the origin of this mismatch. The Z model is built using the same lensing constraints as the L model, including also the spectroscopic redshift of one system at $z_{\rm s}=2.9$. The Z model is also validated by additional two systems of multiply-lensed images, which were identified thanks to the model itself. Therefore, we are confident that the Z model used in the analysis is better constrained than the L model.

To summarize, we find a good agreement between Z and L models in three out of four of the MACS clusters used in this comparison, while for the fourth cluster, we think that the Z model is  better constrained, supporting our results. Therefore, we are confident that out results are robust. In the $\sigma-\theta_{\rm E}$ plane, even the L models do not depart significantly from the relation fitted to the simulations. This is shown in Fig.~\ref{fig:recs_comp}, which shows with sticks and diamonds the new locations of the 4 MACS clusters, when they are modeled using {\rm lenstool}, and their shift with respect to the corresponding Z models. These are overlaid to Fig.~\ref{fig:csthetamacs} for comparison. 

\begin{figure}[lt!]
\begin{center}
  \includegraphics[width=\hsize]{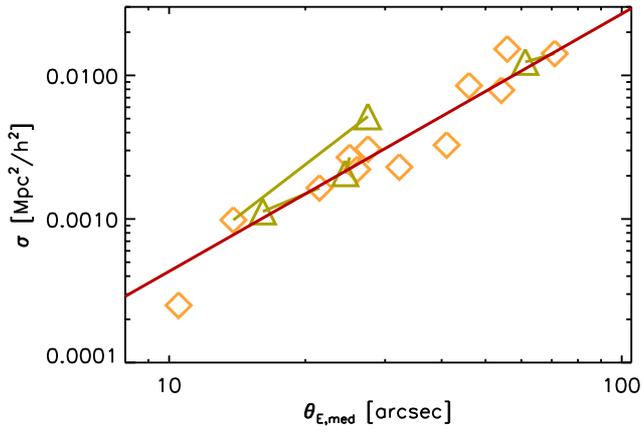}
\end{center}
\caption{The shift of the some of the MACS clusters in the $\sigma-\theta_{\rm E}$, when alternative models obtained with {\tt lenstool} are used to derive the Einstein rings and the lensing cross sections.} 
\label{fig:recs_comp}
\end{figure}
}

\end{document}